%% file: main.tex
\documentclass[10pt,twocolumn,letterpaper]{article}

\usepackage{wacv}              

\input{preamble}

\definecolor{wacvblue}{rgb}{0.21,0.49,0.74}
\usepackage[pagebackref,breaklinks,colorlinks,allcolors=wacvblue]{hyperref}

\def\wacvPaperID{1589} 
\def\confName{WACV}
\def\confYear{2027}

\title{LUMINR: Intensity-conditioned Meta-learned INRs for \linebreak Accelerated Medical Image Registration}

\author{Damian Rynczak$^{1}$, Marek Wodzinski$^{1,2}$ \\
$^{1}$AGH University of Krakow \\
$^{2}$Sano - Centre for Computational Personalised Medicine, International Research Foundation \\
Kraków, Poland \\
{\tt\small drynczak@agh.edu.pl, wodzinski@agh.edu.pl}}

\begin{document}
\maketitle
\input{sec/0_abstract}    
\input{sec/1_intro_wacv}
\input{sec/2_method_wacv}
\input{sec/3_results_wacv}
\input{sec/4_ablation_studies_wacv}
\input{sec/5_discussion_wacv}

\clearpage
{
    \small
    \input{sec/7_acknowneldgement}
    \bibliographystyle{ieeenat_fullname}
    \bibliography{main}
}

\clearpage
\input{sec/6_appendix_wacv}
\end{document}

%% file: preamble.tex
\usepackage{graphicx}
\usepackage{verbatim}
\usepackage{svg}
\usepackage{stfloats}
\usepackage{amsmath,amssymb}
\usepackage{booktabs}
\usepackage{enumitem}



%% file: sec/0_abstract.tex
\begin{abstract}
Implicit neural representations (INRs) have been applied to deformable image registration, where a lightweight network maps a spatial coordinate to a deformation vector. \linebreak Because an INR encodes one specific signal in its parameters, a fresh network must be optimized for every image pair at test time, which is slow. Existing methods reduce this cost by meta-learning an initialization across a population of image pairs, and condition the weights of the INR globally, so the same conditioning applies at every coordinate while the deformation field it initializes varies across the domain. We propose LUMINR, a meta-learning framework that accelerates this per-pair fitting and in which the conditioning is local instead. LUMINR uses an INR that maps a coordinate, together with local intensity and gradient features, to a deformation vector, so the conditioning varies over the domain at the resolution of the field being modelled. Its objective is to find an initialization of the INR parameters from which the network can be fine-tuned to register any unseen pair in only a few optimization iterations. \linebreak We evaluate LUMINR on 3D deformable registration of brain MRI from the OASIS dataset, where local conditioning and the meta-learned initialization prove complementary: early in fitting they reach a markedly higher accuracy than either does alone. Fine-tuned from that initialization, LUMINR matches the accuracy of the same conditioned INR fit from scratch in roughly a tenth of the gradient steps and a tenth of the wall-clock time, while the deformation fields stay smooth with negligible folding.

\end{abstract}

%% file: sec/1_intro_wacv.tex
\section{Introduction}
\label{sec:intro}

Implicit neural representations (INRs) parameterize complex signals as continuous functions, using a neural network to map input coordinates directly to signal values. They are a compact, resolution-independent alternative to discrete representations such as voxel grids. Because of their versatile and continuous nature, INRs have been applied across many areas of medical imaging, including image reconstruction \cite{reed2021,shen2022,wiesner2022}, dataset-level representation learning \cite{friedrich2026}, and predictive tasks such as segmentation \cite{vyas2025}. Deformable image registration is another such task. An INR can represent the coordinate transformation between an image pair directly: a lightweight MLP takes a continuous spatial coordinate as input and returns a deformation vector, which is added back to the coordinate through a residual connection to form the transformation \cite{wolterink2022}.

\begin{figure*}[t]
  \centering
  \includegraphics[width=0.8\linewidth]{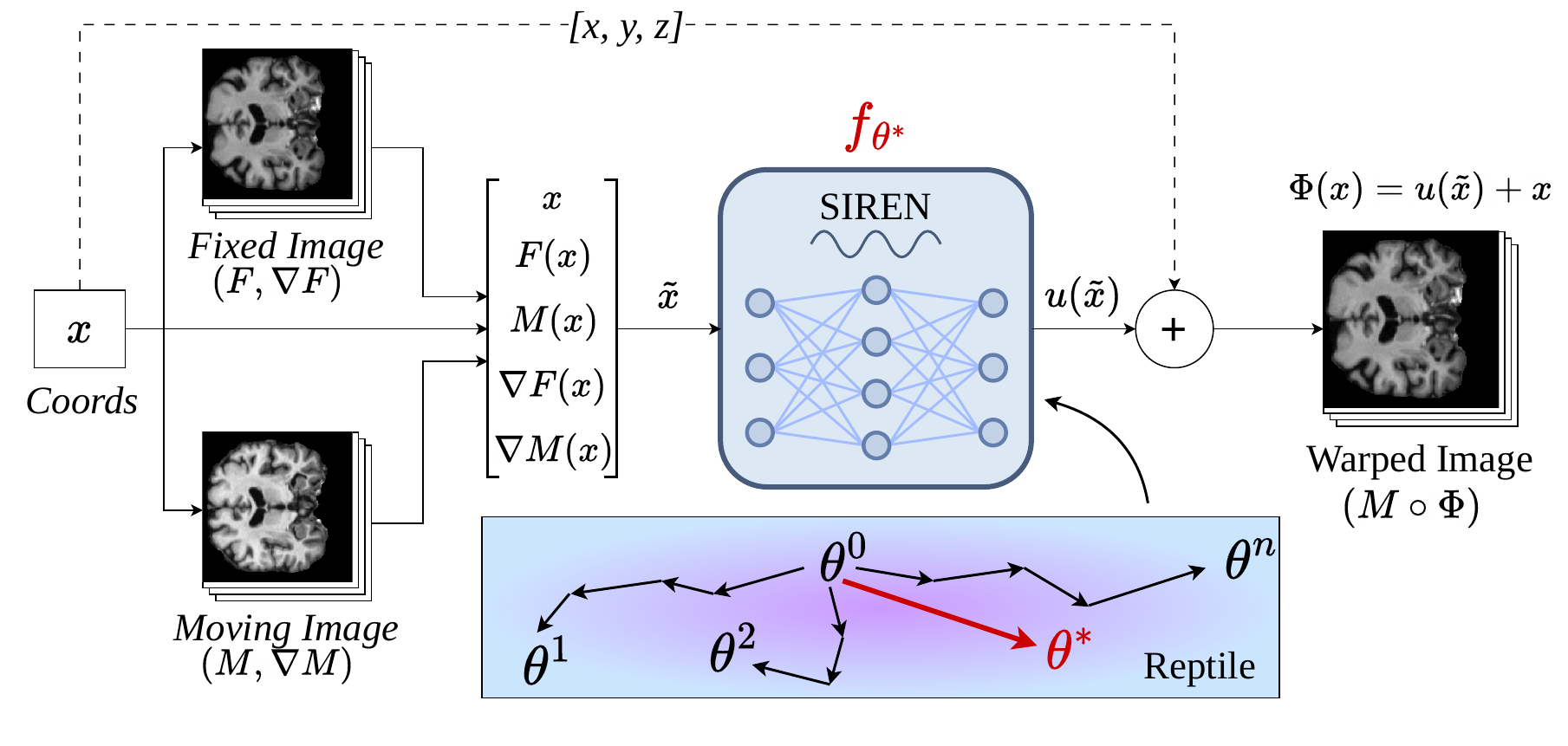}
  \caption{LUMINR uses a multi-layer perceptron (MLP) that implicitly represents the
function $\Phi$, mapping an input feature vector from the moving image to its
corresponding location in the fixed image. Using Reptile meta-learning, LUMINR
finds an initialization $\theta^{*}$ from which $f_\theta$ can be fine-tuned to
the correct correspondence for any unseen image pair in only a few optimization
iterations.}
  \label{fig:LUMINR-diagram-final}
\end{figure*}

\medskip For deformable registration, fitting an INR to each pair has been reported to be more accurate than predicting the transformation with a trained network \cite{wolterink2022}. Such feed-forward encoder-decoder models are trained on large annotated collections and return a deformation field for an unseen pair directly \cite{balakrishnan2019,khan2023,ronneberger2015}; once trained they run faster than conventional iterative methods, but at reduced accuracy \cite{wolterink2022}. An INR is instead fit to the pair at hand \linebreak rather than amortized over a population, so registration remains an optimization problem rather than a prediction problem, and it is this per-pair optimization that accounts for the accuracy.

\medskip These properties come with an important limitation. Because a learned INR is tied so closely to the signal it was fit to, representing a new signal requires fitting a fresh network through iterative optimization. This is slow, and the cost is more acute in deformable image registration, where a separate network must be optimized for every fixed and moving image pair at test time. Several methods aim to reduce this per-instance cost. The shared idea is to replace \linebreak a generic random initialization with one learned across \linebreak a population of signals, so that an INR fits an unseen signal in far fewer gradient steps \cite{dupont2022,tancik2021,vyas2024}. Such learned initializations have been studied for image fitting and reconstruction, and also for deformable image registration, where the initialization is meta-learned over image pairs and then fine-tuned at inference on the pair at hand \cite{baum2023,nichol2018}.

\medskip REINDIR \cite{harten2024} goes one step further and conditions that initialization on the pair to be registered. A convolutional encoder embeds the fixed and moving images into a single vector, and per-layer heads turn it into modulation vectors, one repeated to fill each hidden weight matrix and one multiplying the corresponding bias. These vectors specialize meta-learned template INRs before test-time optimization begins. The network itself still receives a bare spatial \linebreak coordinate as input, and because one vector describes the whole pair, the same conditioning signal applies at every coordinate, while the deformation field it initializes varies across the domain.

\medskip That gap between a conditioning signal held constant over the domain and a field that varies across it limits what the initialization can hold. Vyas et al. \cite{vyas2024} motivate their work by observing that an INR partitions its input coordinate space finely around the spatial gradients of the one signal it was fit to, so that its features do not readily transfer to a second signal even of similar content, and that the partition induced by a meta-learned initialization is closer to random. A bare coordinate says nothing about which pair is being registered, so an initialization shared across a population can encode only what all pairs have in common, and the pair-specific structure is left to gradient descent at test time. Sampling the two intensity fields and their spatial gradients at the query coordinate supplies that structure at the input instead, where it varies from point to point, and the initialization then has only to encode how to use it.

\medskip We propose LUMINR, a meta-learning \cite{finn2017,nichol2018} framework for accelerated medical image registration in which the conditioning is local. The INR maps a spatial coordinate together with the image intensities and gradient features sampled at that coordinate, \linebreak $\tilde{\mathbf{x}} = [\mathbf{x}, F(\mathbf{x}), M(\mathbf{x}), \nabla F(\mathbf{x}), \nabla M(\mathbf{x})]$, to a deformation vector, so the conditioning varies over the domain at the resolution of the field being modelled. LUMINR then seeks an initialization for that INR from which the network can be fine-tuned quickly to any fixed and moving image pair (Fig.~\ref{fig:LUMINR-diagram-final}). An inner loop adapts the parameters to a single sampled pair over several gradient steps by minimizing the registration loss for that pair, and an outer loop moves the shared initialization toward these adapted values \cite{nichol2018}. Repeated across many pairs, this produces an initialization from which fine-tuning converges to an accurate deformation field for an unseen pair in far fewer iterations than fitting from scratch.

\medskip We evaluate LUMINR on 3D deformable registration of brain MRI from the OASIS dataset \cite{hoopes2022,marcus2007}, against coordinate-only INR baselines fitted from random and from meta-learned weights. Fine-tuned from the meta-learned initialization, the network reaches the registration accuracy of the same INR fit from scratch in roughly a tenth of the gradient steps and a tenth of the wall-clock time. The acceleration holds under both a coarse and a fine anatomical segmentation, and it grows as the labels become finer and small boundary displacements matter more. The deformation fields stay smooth throughout fitting, with the fraction of folding voxels below $0.001\%$ at every checkpoint.

\medskip \textbf{Contribution.} Our contributions can be summarized \linebreak as follows:
\begin{itemize}[leftmargin=2em]
  \item We introduce LUMINR, an INR framework for 3D deformable registration whose conditioning is local. Each coordinate is presented with the fixed and moving intensities and their gradients sampled at that point, so the pair-specific signal varies over the domain, whereas existing meta-learned INR registration specializes shared weights from one global embedding \linebreak of the pair.
  \item We meta-learn the initialization of this conditioned network with Reptile, so that an unseen pair is fitted from weights adapted across the training population rather than from random ones.
  \item We show on OASIS brain MRI that local conditioning and the meta-learned initialization are complementary, \linebreak and that together they shorten the test-time fit by roughly an order of magnitude at no cost in accuracy or deformation regularity.
\end{itemize}

%% file: sec/2_method_wacv.tex
\section{Method}
\label{sec:method}

We consider the pairwise registration task for a given dataset $\mathcal{D} = \{(F_j, M_j)\}_{j=1}^{N}$ of $N$ three-dimensional fixed and moving image pairs, where the goal is to recover a spatial transformation that aligns each moving image with its fixed counterpart. Both images of a pair share a normalized domain, $F, M : \Omega \to \mathbb{R}$ with $\Omega \subset [-1, 1]^{3}$. For every image we precompute the spatial gradients $\nabla F$ and $\nabla M$ offline, so that at any continuous coordinate $\mathbf{x} \in \Omega$ we can sample the intensities $F(\mathbf{x}), M(\mathbf{x}) \in \mathbb{R}$ together with their gradients $\nabla F(\mathbf{x}), \nabla M(\mathbf{x}) \in \mathbb{R}^{3}$. Concatenating these with the coordinate yields the input feature vector $\tilde{\mathbf{x}} = [\mathbf{x}, F(\mathbf{x}), M(\mathbf{x}), \nabla F(\mathbf{x}), \nabla M(\mathbf{x})] \in \mathbb{R}^{d}$ ($d = 11$), so the network has access to local appearance context as well as position.

\medskip LUMINR represents the deformation with an $L$-layer MLP $f_\theta : \mathbb{R}^{d} \to \mathbb{R}^{3}$ that maps $\tilde{\mathbf{x}}$ to a displacement $\mathbf{u}(\tilde{\mathbf{x}})$. A residual connection adds the original coordinate back to form the transformation $\Phi(\mathbf{x}) = \mathbf{u}(\tilde{\mathbf{x}}) + \mathbf{x}$. The warped moving image is obtained by sampling $M$ at the transformed coordinates $\Phi(\mathbf{x})$, with subvoxel intensities resolved through trilinear interpolation \cite{wolterink2022}.
We seek the transformation that maps each fixed-image coordinate to its anatomically corresponding location in the moving image, so that $(M \circ \Phi)(\mathbf{x}) = F(\mathbf{x})$ for all $\mathbf{x} \in \Omega$. LUMINR's objective is to find an initialization $\theta^{*}$ of the INR parameters from which $f_\theta(\cdot)$ can be fine-tuned to satisfy this correspondence for any unseen pair in only a few optimization iterations \cite{nichol2018,wolterink2022}. In the following Sec. \ref{sec:meta-reg-desc}, we describe how Reptile meta-learning is used to obtain this initialization.

\begin{figure*}[t]
  \centering
  \includegraphics[width=\textwidth]{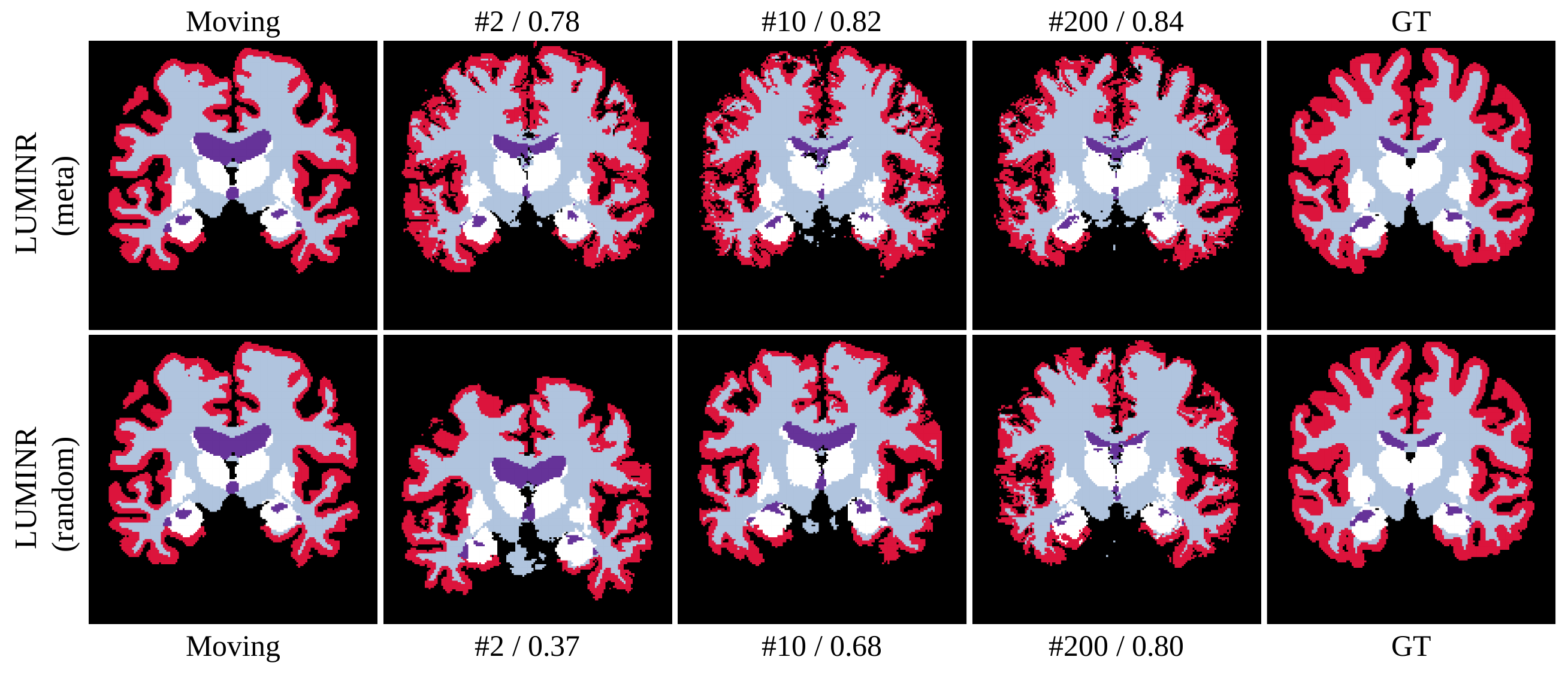}
  \caption{Comparison of segmentation performance for the LUMINR (meta) and LUMINR (random) models ($\alpha$ = 0.1) across 4 classes. We report the mean Dice score at each optimization iteration for a single pair of the MRI volumes. The meta-learned initialization leads to rapid convergence and achieves reasonable segmentation (Dice = 0.82) after 10 iterations.}
  \label{fig:masks-4classes}
\end{figure*}

\subsection{LUMINR: Initialization for Accelerated Medical Image Registration}
\label{sec:meta-reg-desc}

LUMINR uses the Reptile meta-learning strategy \cite{nichol2018} to learn an initialization $\theta^{*}$ from which the network adapts quickly to a new fixed and moving image pair. We chose Reptile over MAML \cite{finn2017} because MAML differentiates through the inner-loop update and so requires the Hessian, whose similarity term reaches the parameters through the trilinear resampler that warps the moving image. Trilinear interpolation is only $C^0$: within a voxel cell the interpolant is multilinear, so its pure second derivatives vanish identically, and the mixed terms that remain are in general discontinuous across cell faces. At those faces the true second derivative is a singular measure that automatic differentiation cannot access. This path therefore contributes degenerate curvature. Reptile's update depends only on the fine-tuned parameters, so it needs no second derivatives, and its memory cost does not grow with the number of inner steps.

\medskip Training is a nested optimization with an \textit{inner} and an \textit{outer} routine. The inner routine treats each pair as a separate task: given a sampled pair $(F_j, M_j)$, it adapts the current parameters $\theta^{t}$ by running $T_i$ Adam steps \cite{kingma2014} on the registration loss of $f_{\theta}(\cdot)$, which yields the adapted weights $\tilde{\theta}_j$ \cite{nichol2018,wolterink2022}. The outer routine begins from random values $\theta^{0}$ at $t = 0$ and runs for $T_o$ meta-iterations. At each iteration it calls the inner routine on a freshly sampled pair and moves the shared parameters toward the adapted weights, so that after training a few gradient steps on an unseen pair are enough to reach a low registration loss.

\medskip Deformable image registration seeks the transformation ${\Phi}$ that best aligns the warped moving image with the fixed image while remaining physically plausible:

\begin{equation}
{\Phi} \;=\; \arg\min_{\Phi}\; \mathcal{L}_{\text{data}}\big(M\circ\Phi,\, F\big) \;+\; \alpha\,\mathcal{L}_{\text{reg}}(\Phi),    
\end{equation}

where $\alpha$ weights the regularization against the data term. At each inner iteration $t$ we form this objective for the sampled pair $j$ and optimize the SIREN \cite{sitzmann2020} network $f_{\theta^t}$, which predicts the displacement $\mathbf{u}(\tilde{\mathbf{x}}) = f_{\theta^t}(\tilde{\mathbf{x}})$ from the enriched input $\tilde{\mathbf{x}}$. A residual connection then gives the transformation $\Phi_{\theta^t}(\mathbf{x}) = \mathbf{u}(\tilde{\mathbf{x}}) + \mathbf{x}$. We optimize the network using the following loss term:

\begin{equation}
\begin{split}
\mathcal{L}_{\text{inner}}\big(F_j, M_j\big) \;=\; {}
  &-\,\mathcal{L}_{\text{NCC}}\big(M_j \circ \Phi_{\theta},\, F_j\big) \\
  &+\,\alpha\,\mathcal{L}_{\text{reg}}\big(\Phi_{\theta}\big),
\end{split}
\label{eq:inner_loss}
\end{equation}

\medskip The data term $\mathcal{L}_{\text{NCC}}$ is the normalized cross-correlation between the fixed intensities and the warped moving intensities, evaluated on a batch of $N_s = 10{,}000$ coordinates sampled from the image domain at each iteration. Because the warped coordinates generally fall off the voxel grid, the moving image is sampled by trilinear interpolation, which keeps the data term differentiable end to end \cite{wolterink2022}. For the regularization term $\mathcal{L}_{\text{reg}}$ we use the bending-energy penalty \cite{rueckert1999}, which enforces smoothness on the deformation vector field by keeping its second-order spatial derivatives small throughout the domain. This penalty suits SIREN well: its periodic activations can be differentiated more than once, so the required second derivatives are available analytically, whereas they would vanish for a ReLU network  \cite{wolterink2022}.

\medskip A single inner step ($T_i = 1$) reduces to plain Stochastic Gradient Descent (SGD) on the expected loss and yields no useful initialization; only $T_i > 1$ introduces the higher-order term that drives within-task generalization, and \linebreak a larger $T_i$ makes $\tilde{\theta}_j$ a closer approximation of pair $j$'s task-optimal solution. We therefore use a moderately large number of inner steps (e.g., $T_i$ between 100 and 200). The outer routine then treats the displacement $(\tilde{\theta}_j - \theta^{t})$ as \linebreak a pseudo-gradient and moves the initialization towards the \linebreak adapted weights:

\begin{equation}
\theta^{t+1} \;\leftarrow\; \theta^t + \epsilon\,\big[\tilde{\theta}_j - \theta^t\big],    
\end{equation}

where $\epsilon$ is the outer step size. After $T_o$ outer steps we obtain the meta-learned initialization $\theta^{*} = \theta^{T_o}$.

\medskip At inference we are given an unseen fixed-moving pair $(F, M)$. As in classical INR approach, we fit the network to this single pair by optimizing the same objective (\ref{eq:inner_loss}) used in the inner loop for $T_f$ iterations. The only difference is the starting point. Instead of a random initialization, we initialize $f_\theta(\cdot)$ with meta-learned parameters $\theta^{*}$. Once fitted, a single feed-forward pass recovers the deformation $\Phi_{\theta^{}}(\mathbf{x}) = f_{\theta^{}}(\tilde{\mathbf{x}}) + \mathbf{x}$ at any coordinate $\mathbf{x} \in \Omega$. We then warp the moving image with trilinear interpolation, $\big(M \circ \Phi_{\theta^{}}\big)(\mathbf{x})$, to obtain the registered image aligned to the fixed image.

\begin{figure*}[t]
  \centering
  \includegraphics[width=\textwidth]{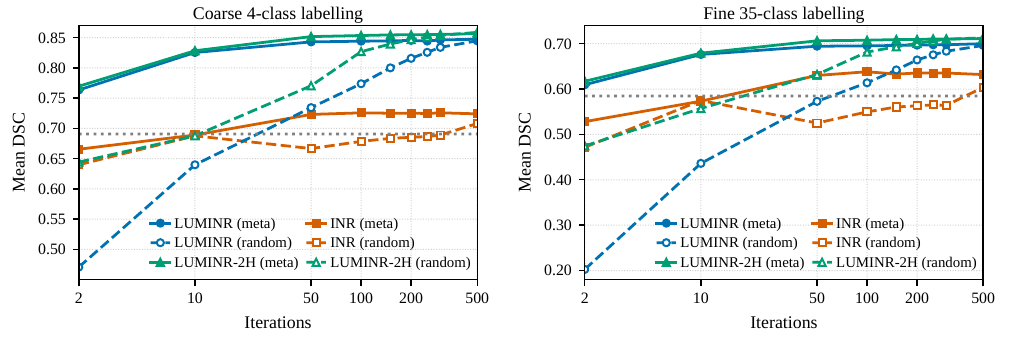}
  \caption{Mean Dice score over all test pairs at each fitting iteration ($\alpha=0.1$), on the coarse 4-class (left) and the fine 35-class (right) labelling, background included; the grey dotted line is the pre-registration, affine-only Dice ($0.691$ and $0.584$).  LUMINR (meta) matches at 50 steps the Dice that LUMINR (random) reaches only at 500, and by two steps it already exceeds every point of both INR curves on DSC$_4$; \linebreak on DSC$_{35}$ two steps match the INR baseline at 500 steps and ten steps clear INR (meta).}
  \label{fig:dsc-init}
\end{figure*}

\subsection{Experimental setup}
\label{sec:exper-setup}

\textbf{Dataset and preprocessing.} We evaluated LUMINR on aligned 3D brain MRI volumes from the OASIS dataset \cite{hering2022,hoopes2022,marcus2007}. We took 414 T1-weighted scans and split them randomly into 280 training, 34 validation, and 100 test volumes. Each volume comes with segmentations at two granularities: a coarse set of 4 labels (cortex, subcortical gray matter, white matter, and CSF) and a fine set of 35 labels for individual anatomical regions. We normalized all volumes to the intensity range [0,1] and kept the original resolution of 160\;x\;192\;x\;224. In addition, we computed the image gradients with a Sobel operator and used the raw gradient magnitudes to build input feature vector. 

\medskip OASIS contains unpaired images, so we built registration pairs explicitly. For training, each subject is paired once with every other subject, which gives every unordered fixed-moving combination. At each outer optimization step we drew a new batch of 300 pairs at random from this pool, and for each pair we evaluated the registration loss on a random minibatch of 10,000 spatial coordinates rather than the full image grid. For evaluation we instead pair the 100 test volumes consecutively, which gives 99 test pairs.

\medskip \textbf{Implementation.} For all experiments we used a SIREN INR with $L = 5$ hidden layers, width $h = 512$, and $\omega_0 = 30$. We optimized the network with Adam \cite{kingma2014}, using \linebreak a learning rate of $1 \times 10^{-4}$ for the inner loop and $1 \times 10^{-3}$ for the outer meta-learning loop. We set $T_i = 100$ inner steps and ran the outer loop for $T_o = 100$ iterations, applying an $\alpha$-weighted bending-energy regularizer with $\alpha = 0.1$. We validated every 5 outer steps using 50 inner steps. We also evaluated a second input configuration (LUMINR-2H), in which two separate input layers process coordinates and intensity features independently and their outputs are concatenated before the next layer. Aside from this split input stage, LUMINR2H uses the same architecture \linebreak and optimization settings.

\medskip At inference time, we initialized $f_\theta$ with the learned meta-parameters $\theta^*$ and fit each unseen scan for 500 iterations with Adam at a learning rate of $1 \times 10^{-4}$ and the bending-energy regularizer set to $\alpha=0.1$. All fits used coordinate batches of 10,000 points. \linebreak We also swept $\alpha \in \{0.01,0.1,1.0\}$ for both the meta-learning and test-time fitting stages to measure its effect on registration quality.

\medskip \textbf{Baseline.} We compared our method against a baseline INR that receives only the spatial coordinate as input, without the sampled intensity and gradient features, so that $f_\theta : \mathbb{R}^{3} \to \mathbb{R}^{3}$. Apart from the input, this baseline is identical to LUMINR: the same SIREN backbone, the same registration objective (\ref{eq:inner_loss}), and the same optimization settings described above. The comparison therefore isolates the contribution of the enriched input representation.

\medskip We ran both models under two initialization regimes. In the first, the network starts from randomly initialized weights following the standard SIREN scheme, and is fit to each unseen test pair for $T_f = 500$ iterations with the Adam optimizer, as in the classical INR approach. In the second, the same fitting procedure starts from the meta-learned parameters $\theta^{*}$. The meta-learning stage of the baseline used the settings reported above for LUMINR. Together the four configurations separate the effect of the intensity features from the effect of the initialization, and show how each of them affects registration speed and accuracy.

\medskip \textbf{Grid-searched baseline.} We also compared LUMINR against a coordinate-only INR whose hyperparameters we selected by grid search. It is a SIREN with two layers and 256 hidden features, optimized at a learning rate of $6.6 \times 10^{-5}$. We evaluated this baseline in the same two regimes, from scratch and from a meta-learned initialization obtained with the same meta-learning parameters as described above. We describe the search space and the selection procedure in detail in the supplementary material (Sec.~\ref{sec:supp-grid-search}).

%% file: sec/3_results_wacv.tex
\section{Results}
\label{sec-results}
We report mean Dice score, cumulative fitting time, and the fraction of folding voxels over the 99 OASIS test pairs, computed at full resolution and tracked from iteration 2 to iteration 500 at $\alpha=0.1$. Our objective is to accelerate INR fitting on an unseen medical image pair, so we cap fitting at 500 iterations and ask how local intensity-feature conditioning and the meta-learned weight initialization affect the early stages of optimization rather than the converged solution. We compare six main configurations: the enriched-input model (LUMINR) and two coordinate-only baselines, one sharing the same SIREN backbone (INR) and one whose hyperparameters we selected by grid search (INR$_{\mathrm{GS}}$), each fitted from randomly initialized weights (denoted \textit{random}) and from the meta-learned parameters \linebreak (denoted \textit{meta}). Every test volume carries two segmentations, one over 4 classes and background and one over 35 classes and background, and we report Dice score for both.

\begin{table*}[t]
    \centering
    \caption{Mean Dice score, cumulative fitting time, and fraction of folding voxels across fitting iterations for the LUMINR (meta) and LUMINR (random) models. DSC$_4$ and DSC$_{35}$ are Dice on the coarse (4-class) and fine (35-class) labellings respectively; fitting time and folding are independent of the labelling and are therefore reported once. Each entry is the mean over all test pairs with its 95\% confidence interval in brackets. Both models are the same architecture: the left one is fitted from the meta-learned initialization, the right one from \linebreak a random initialization.}
    \label{tab:mri-luminr-meta-vs-scratch-ci}
    \footnotesize
    \newcommand{\itc}[1]{\raisebox{0.5\ht\strutbox}[0pt][0pt]{#1}}
    \setlength{\tabcolsep}{4pt}
    \resizebox{\textwidth}{!}{%
    \begin{tabular}{@{}ccccccccc@{}}
    \toprule
     & \multicolumn{4}{c}{LUMINR (meta)} & \multicolumn{4}{c}{LUMINR (random)} \\
    \cmidrule(lr){2-5} \cmidrule(lr){6-9}
    Iter. & DSC$_4$ & DSC$_{35}$ & $t$ & det($J$)$<$0\,$\times10^{-5}$ & DSC$_4$ & DSC$_{35}$ & $t$ & det($J$)$<$0\,$\times10^{-5}$ \\
          & [-] & [-] & [s] & [\%] & [-] & [-] & [s] & [\%] \\
    \midrule
    \itc{2}   & \shortstack{0.764 \\ {\footnotesize [0.756, 0.771]}} & \shortstack{0.610 \\ {\footnotesize [0.598, 0.620]}} & \shortstack{0.27 \\ {\footnotesize [0.26, 0.27]}} & \shortstack{0.26 \\ {\footnotesize [0.06, 0.56]}} & \shortstack{0.471 \\ {\footnotesize [0.460, 0.481]}} & \shortstack{0.202 \\ {\footnotesize [0.189, 0.217]}} & \shortstack{0.28 \\ {\footnotesize [0.26, 0.29]}} & \shortstack{$0$ \\ {\footnotesize [0, 0]}} \\
    \itc{10}  & \shortstack{0.825 \\ {\footnotesize [0.819, 0.832]}} & \shortstack{0.676 \\ {\footnotesize [0.664, 0.688]}} & \shortstack{1.32 \\ {\footnotesize [1.32, 1.33]}} & \shortstack{0.94 \\ {\footnotesize [0.00, 2.73]}} & \shortstack{0.640 \\ {\footnotesize [0.632, 0.647]}} & \shortstack{0.436 \\ {\footnotesize [0.421, 0.452]}} & \shortstack{1.31 \\ {\footnotesize [1.30, 1.33]}} & \shortstack{$0$ \\ {\footnotesize [0, 0]}} \\
    \itc{50}  & \shortstack{0.843 \\ {\footnotesize [0.837, 0.849]}} & \shortstack{0.694 \\ {\footnotesize [0.683, 0.706]}} & \shortstack{6.55 \\ {\footnotesize [6.55, 6.56]}} & \shortstack{1.41 \\ {\footnotesize [0.46, 2.61]}} & \shortstack{0.734 \\ {\footnotesize [0.728, 0.740]}} & \shortstack{0.573 \\ {\footnotesize [0.562, 0.583]}} & \shortstack{6.50 \\ {\footnotesize [6.48, 6.52]}} & \shortstack{0.32 \\ {\footnotesize [0.00, 0.95]}} \\
    \itc{100} & \shortstack{0.844 \\ {\footnotesize [0.838, 0.850]}} & \shortstack{0.695 \\ {\footnotesize [0.683, 0.706]}} & \shortstack{13.12 \\ {\footnotesize [13.09, 13.19]}} & \shortstack{2.07 \\ {\footnotesize [0.48, 4.68]}} & \shortstack{0.774 \\ {\footnotesize [0.767, 0.780]}} & \shortstack{0.614 \\ {\footnotesize [0.602, 0.625]}} & \shortstack{12.98 \\ {\footnotesize [12.96, 13.00]}} & \shortstack{13.42 \\ {\footnotesize [0.13, 39.06]}} \\
    \itc{150} & \shortstack{0.845 \\ {\footnotesize [0.838, 0.850]}} & \shortstack{0.696 \\ {\footnotesize [0.685, 0.707]}} & \shortstack{19.67 \\ {\footnotesize [19.63, 19.73]}} & \shortstack{1.07 \\ {\footnotesize [0.26, 2.16]}} & \shortstack{0.800 \\ {\footnotesize [0.793, 0.806]}} & \shortstack{0.642 \\ {\footnotesize [0.631, 0.653]}} & \shortstack{19.45 \\ {\footnotesize [19.43, 19.47]}} & \shortstack{14.50 \\ {\footnotesize [1.37, 33.84]}} \\
    \itc{200} & \shortstack{0.845 \\ {\footnotesize [0.839, 0.851]}} & \shortstack{0.697 \\ {\footnotesize [0.685, 0.708]}} & \shortstack{26.21 \\ {\footnotesize [26.17, 26.28]}} & \shortstack{1.20 \\ {\footnotesize [0.46, 2.13]}} & \shortstack{0.816 \\ {\footnotesize [0.809, 0.822]}} & \shortstack{0.664 \\ {\footnotesize [0.653, 0.675]}} & \shortstack{25.92 \\ {\footnotesize [25.90, 25.94]}} & \shortstack{5.67 \\ {\footnotesize [0.10, 15.50]}} \\
    \itc{250} & \shortstack{0.844 \\ {\footnotesize [0.838, 0.850]}} & \shortstack{0.698 \\ {\footnotesize [0.686, 0.709]}} & \shortstack{32.75 \\ {\footnotesize [32.71, 32.82]}} & \shortstack{1.14 \\ {\footnotesize [0.44, 2.01]}} & \shortstack{0.826 \\ {\footnotesize [0.819, 0.832]}} & \shortstack{0.676 \\ {\footnotesize [0.665, 0.686]}} & \shortstack{32.39 \\ {\footnotesize [32.38, 32.42]}} & \shortstack{3.33 \\ {\footnotesize [0.87, 6.61]}} \\
    \itc{300} & \shortstack{0.846 \\ {\footnotesize [0.840, 0.852]}} & \shortstack{0.698 \\ {\footnotesize [0.686, 0.709]}} & \shortstack{39.30 \\ {\footnotesize [39.25, 39.38]}} & \shortstack{1.17 \\ {\footnotesize [0.57, 1.91]}} & \shortstack{0.834 \\ {\footnotesize [0.828, 0.840]}} & \shortstack{0.683 \\ {\footnotesize [0.672, 0.694]}} & \shortstack{38.87 \\ {\footnotesize [38.85, 38.89]}} & \shortstack{3.58 \\ {\footnotesize [0.70, 7.40]}} \\
    \itc{500} & \shortstack{0.847 \\ {\footnotesize [0.842, 0.853]}} & \shortstack{0.700 \\ {\footnotesize [0.689, 0.711]}} & \shortstack{65.53 \\ {\footnotesize [65.44, 65.66]}} & \shortstack{1.10 \\ {\footnotesize [0.47, 1.86]}} & \shortstack{0.845 \\ {\footnotesize [0.838, 0.851]}} & \shortstack{0.696 \\ {\footnotesize [0.685, 0.707]}} & \shortstack{64.79 \\ {\footnotesize [64.75, 64.85]}} & \shortstack{3.08 \\ {\footnotesize [0.25, 7.85]}} \\
    \midrule
    $\Delta$DSC$_{4,\,(50\to500)}$ & \multicolumn{4}{c}{$\textbf{+0.5\%}$} & \multicolumn{4}{c}{$+15.1\%$} \\
    $\Delta$DSC$_{35,\,(50\to500)}$ & \multicolumn{4}{c}{$\textbf{+0.9\%}$} & \multicolumn{4}{c}{$+21.5\%$} \\
    \bottomrule
    \end{tabular}}
    \end{table*}

\subsection{Effect of the enriched input representation}
\label{sec-results-conditioning}

\begin{figure*}[t]
  \centering
  \includegraphics[width=\textwidth]{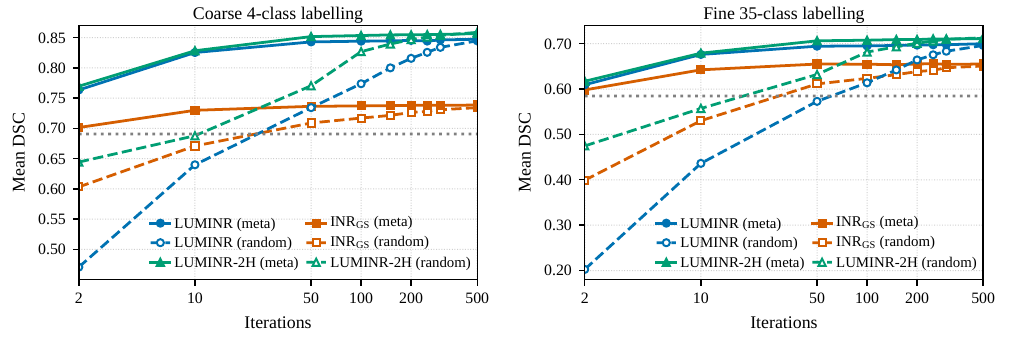}
  \caption{Mean Dice score over all test pairs at each evaluated fitting iteration ($\alpha=0.1$), on the coarse 4-class (left) and the fine 35-class (right) labelling, background included; the grey dotted line is the pre-registration, affine-only Dice ($0.691$ and $0.584$).  At 50 steps LUMINR (meta) comes within $0.002$ of the Dice that LUMINR (random) reaches only at 500, and by two steps it already exceeds every point of both INR$_{\mathrm{GS}}$ curves on DSC$_4$; on DSC$_{35}$ two steps match INR$_{\mathrm{GS}}$ at 50 steps and ten steps clear both curves.}
  \label{fig:dsc-init-gs}
\end{figure*}

\textbf{Fitting from scratch.} We began by analyzing the effect of the enriched input representation on network fitting. We observed that conditioning on local intensity features improves registration accuracy beyond the initial fitting iterations. LUMINR (random) attains a higher mean Dice score than both coordinate-only baselines from iteration 50 onwards on the coarse labelling and from iteration 150 onwards on the fine one, and this ordering is preserved at every subsequent checkpoint (Fig.~\ref{fig:dsc-init}, Fig.~\ref{fig:dsc-init-gs}). Over the interval from iteration 50 to iteration 500 the conditioned model gains more Dice score than either INR or INR$_{\mathrm{GS}}$ on both labellings, its margin over each baseline attains its maximum in the second half of the fitting schedule, and it remains superior at the terminal iteration. We attribute this to the input representation: because the fixed and moving intensities and their gradients are sampled at the query coordinate, the network observes the local appearance mismatch that the displacement must resolve rather than inferring it from position alone.

\medskip \textbf{Fitting from the meta-learned initialization.} Since this advantage needed several tens of iterations to appear, we next asked whether a better starting point removes the delay. Starting from the meta-learned parameters, LUMINR (meta) attains a higher mean Dice score than both INR (meta) and INR$_{\mathrm{GS}}$ (meta) at every evaluated iteration and on both labellings (Fig.~\ref{fig:dsc-init}, Fig.~\ref{fig:dsc-init-gs}). No early crossover of the kind observed under random initialization occurs: the advantage is present at the second iteration and persists after all models have plateaued. It is more pronounced on the coarse labelling, where the margin over either baseline exceeds the corresponding margin on the fine labelling at every checkpoint. Conditioning and the meta-learned initialization are therefore complementary: intensity features that improve accuracy only from iteration 50 onwards on the coarse labelling and from iteration 150 onwards on the fine one under random initialization contribute from the second iteration when fitting starts from $\theta^{*}$.

\medskip \textbf{Interaction between conditioning and initialization.} We quantified this complementarity by reversing the comparison: how much the meta-learned initialization is worth depends on what the network sees at its input. Averaged over the first fifty fitting iterations, starting from $\theta^{*}$ gains LUMINR $+0.196$ DSC$_4$ and $+0.256$ DSC$_{35}$, against $+0.061$ and $+0.118$ for INR$_{\mathrm{GS}}$ and $+0.027$ and $+0.053$ for INR. The separation is widest at the second iteration, where the initialization is worth $+0.293$ DSC$_4$ and $+0.408$ DSC$_{35}$ to the conditioned model, two to three times what it gives INR$_{\mathrm{GS}}$ and seven to twelve times what it gives INR.

\medskip The same ordering holds for how long the gain survives. Without intensity features the head start is also short-lived. INR (meta) and INR from scratch meet at the tenth iteration, $0.689$ against $0.688$ on DSC$_4$ and $0.573$ against $0.575$ on DSC$_{35}$, so ten gradient steps erase it. Conditioning also raises the accuracy the initialization converges to: from $\theta^{*}$ the coordinate-only models plateau near $0.73$ DSC$_4$, a level LUMINR (meta) passes at its second iteration ($0.764$ after $0.27$\,s) before settling near $0.85$. For a coordinate-only INR, meta-learning mainly moves a low plateau earlier in the schedule. Combined with local conditioning, it reaches a markedly higher one within a few gradient steps.

\subsection{Accelerated medical image registration}
\label{sec-results-acceleration}

We then measured what the combination of the two saves in practice, by reading off the iteration count and the wall-clock time at which each model reaches a matched Dice.

\medskip \textbf{Coarse labelling.} For the 4-classes and background setting (Table~\ref{tab:mri-luminr-meta-vs-scratch-ci}), LUMINR (meta) reaches a Dice of $0.843$ after 50 gradient steps ($6.6$\,s), a score the same network fitted from random weights attains only after the full 500 steps ($0.845$, $64.8$\,s); the meta-learned initialization matches the same registration quality with roughly a tenth of the iterations and a tenth of the wall-clock time. The gap is largest early in fitting: within its first ten steps ($1.3$\,s) the meta-initialized fit is already level with what the random start needs about 250 steps ($32.4$\,s) to reach, and a two-step fit ($0.3$\,s) already surpasses the 500-step plateaus of both coordinate-only baselines ($0.734$ at $31.8$\,s for INR$_{\mathrm{GS}}$, $0.708$ at $91.0$\,s for INR) as well as the random start at 50 steps.

\medskip \textbf{Fine labelling.} The 35-classes-and-background setting follows the same pattern. Within 50 steps ($6.6$\,s) meta-learned LUMINR reaches the Dice that LUMINR (random) recovers only at 500 steps ($64.8$\,s), again about a tenfold reduction in both iterations and time. The early advantage is larger here: a two-step fit already exceeds what LUMINR (random) reaches after 50 steps, and a ten-step fit ($1.3$\,s) is above the plateau INR$_{\mathrm{GS}}$ (random) reaches after its full 500-step schedule ($31.8$\,s). The finer segmentation penalizes small boundary displacements, so fitting from scratch needs many more steps to resolve them, and the acceleration from the meta-learned initialization is larger.

\medskip \textbf{Folding.} We finally verified that the acceleration does not come at the cost of deformation regularity. Across all six configurations the fraction of voxels with a negative Jacobian determinant stays below $0.001\%$ at every checkpoint; because folding is a property of the deformation field, it is identical for the 4- and 35-class evaluations. Neither coordinate-only baseline folds a single voxel in either regime. Only the conditioned models fold at all, and from iteration 100 onwards the meta-learned initialization is the better behaved of the two: its peak folding fraction, $2.07\times10^{-5}\%$ at iteration 100, is roughly seven times lower than the peak of the same architecture fitted from scratch, $14.50\times10^{-5}\%$ at iteration 150. As Fig.~\ref{fig:masks-4classes} illustrates, LUMINR enables the generation of a viable segmentation mask close to the ground truth in just 2 - 10 \linebreak fitting iterations. 

\medskip Supplementary quantitative and qualitative results are presented in Appendix~\ref{sec:quant} and Appendix~\ref{sec:qual}, respectively.


%% file: sec/4_ablation_studies_wacv.tex
\subsection{Ablation studies}
\label{sec-abolation-studies}

In this section, we present ablation studies on two key design choices of LUMINR: the number of enriched input layers, which affects how coordinate and intensity features are processed, and the regularization strength $\alpha$, which controls the trade-off between segmentation accuracy and folding.

\medskip \textbf{Ablation studies.} First, we conducted an ablation study on the effect of adding a second enriched input layer to the meta-learned initialization, comparing LUMINR and LUMINR-2H. The second layer yields only a marginal Dice gain at nearly identical fitting time and negligible folding, so we adopt single input head LUMINR as our primary model (Fig.~\ref{fig:dsc-init}, Fig.~\ref{fig:dsc-init-gs})

\begin{figure*}[t]
  \centering
  \includegraphics[width=\textwidth]{"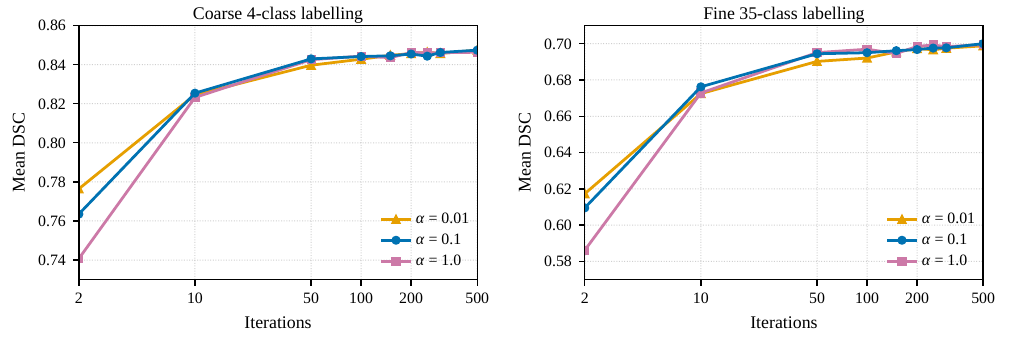"}
  \caption{Mean Dice over all test pairs at each fitting iteration for LUMINR at three regularization strengths ($\alpha \in \{0.01, 0.1, 1.0\}$); 4-class and 35-class and background labelling. The choice of $\alpha$ barely affects the converged score: the settings agree by 50 steps and plateau together at 500, differing only in the first few iterations where weaker regularization ($\alpha=0.01$) starts slightly higher.}
  \label{fig:dsc-alpha-4classes}
\end{figure*}

\medskip \textbf{Regularization strength.} We also conducted an ablation study on how different regularization strengths \linebreak $\alpha$ affect the Dice scores of LUMINR (meta), considering \linebreak $\alpha \in \{0.01, 0.1, 1.0\}$ on both the 4-class and 35-class segmentations (Fig.~\ref{fig:dsc-alpha-4classes}). Overall, the choice of $\alpha$ has only \linebreak a limited effect on the final Dice score, with the three settings yielding comparable results on both tasks. Minor differences are mostly visible in the first few iterations, where weaker regularization tends to start from a slightly higher Dice, though this difference largely disappears as fitting progresses and the three settings become broadly similar. Given that the Dice score remains fairly stable across \linebreak a wide range of $\alpha$, we adopt $\alpha=0.1$ as our final regularization value, as it maintains this accuracy while keeping \linebreak folding low.

%% file: sec/5_discussion_wacv.tex
\section{Discussion}

We have shown that conditioning the INR locally and meta-learning its initialization are complementary, together \linebreak accelerating INR-based deformable image registration. Fine-tuned from the meta-learned initialization, the network reaches the accuracy of the same INR fit from scratch in roughly a tenth of the gradient steps and a tenth of the wall-clock time, under both a coarse and a fine segmentation, while folding stays negligible. This is a step toward INR-based registration that is fast enough for interactive clinical use. The remaining cost is dominated by the regularizer, not the data term. Total fitting time scales with the number of backward passes per epoch, and regularizers that we are using are computed by analytically differentiating the transformation $\Phi$ with respect to the input \linebreak coordinates through autograd, rather than by finite differences on a grid \cite{wolterink2022}. Within the regularization term, the bending energy penalty is the most expensive component, because it requires second-order derivatives of the network output. Reducing this cost is what would bring INR-based registration close to real time and make it usable during live clinical procedures.

\medskip Our method is not immune to folding. The enriched-input models introduce a small, non-zero fraction of folded voxels, and the meta-learned initialization inherits this rather than removing it. This motivates further work on INR regularization. One direction is to design INRs that need less regularization in the first place, for example by addressing the frequency behavior that makes SIREN-based fields prone to high-frequency artifacts.

\medskip A second direction concerns how the INR is conditioned on the image pair. Here we condition on local intensity and gradient features at each coordinate. The present work tests this local-conditioning strategy for meta-learned INR registration rather than benchmarking it against established registration methods; we leave that comparison to future work. These hand-chosen features could instead be replaced by a learned feature vector from a foundation model, which would carry more context about the anatomy being registered. Taken further, the same idea points toward an INR-based foundation model for image registration, in which \linebreak a single meta-learned representation adapts to a wide range of image pairs and, we expect, surpasses the registration performance of current methods.

%% file: sec/7_acknowneldgement.tex
\section*{Acknowledgements}
This work was supported by the National Science Centre, Poland, under Grant ``MultiGeoMed'' No. 2024/55/D/ST6/02081. We gratefully acknowledge Polish high-performance computing infrastructure PLGrid (HPC Center: ACK Cyfronet AGH) for providing computer facilities and support within computational grants no. PLG/2026/019392 \linebreak and PLG/2025/018770.

%% file: sec/6_appendix_wacv.tex
\appendix

\begin{table*}[t]
    \centering
    \caption{Mean Dice score, cumulative fitting time, and fraction of folding voxels across fitting iterations for the baseline INR (meta) and INR (random) models. DSC$_4$ and DSC$_{35}$ are Dice on the coarse (4-class) and fine (35-class) labellings respectively; fitting time and folding are independent of the labelling and are therefore reported once. Each entry is the mean over all test pairs with its 95\% confidence interval in brackets. Both models are the same architecture: the left one is fitted from the meta-learned initialization, the right one from \linebreak a random initialization.}
    \label{tab:mri-inr-meta-vs-scratch-ci}
    \footnotesize
    \newcommand{\itc}[1]{\raisebox{0.5\ht\strutbox}[0pt][0pt]{#1}}
    \setlength{\tabcolsep}{4pt}
    \resizebox{\textwidth}{!}{%
    \begin{tabular}{@{}ccccccccc@{}}
    \toprule
     & \multicolumn{4}{c}{INR (meta)} & \multicolumn{4}{c}{INR (random)} \\
    \cmidrule(lr){2-5} \cmidrule(lr){6-9}
    Iter. & DSC$_4$ & DSC$_{35}$ & $t$ & det($J$)$<$0\,$\times10^{-5}$ & DSC$_4$ & DSC$_{35}$ & $t$ & det($J$)$<$0\,$\times10^{-5}$ \\
          & [-] & [-] & [s] & [\%] & [-] & [-] & [s] & [\%] \\
    \midrule
    \itc{2}   & \shortstack{0.665 \\ {\footnotesize [0.657, 0.674]}} & \shortstack{0.528 \\ {\footnotesize [0.512, 0.544]}} & \shortstack{0.35 \\ {\footnotesize [0.32, 0.39]}} & \shortstack{$0$ \\ {\footnotesize [0, 0]}} & \shortstack{0.640 \\ {\footnotesize [0.631, 0.648]}} & \shortstack{0.471 \\ {\footnotesize [0.456, 0.487]}} & \shortstack{0.75 \\ {\footnotesize [0.59, 0.95]}} & \shortstack{$0$ \\ {\footnotesize [0, 0]}} \\
    \itc{10}  & \shortstack{0.689 \\ {\footnotesize [0.681, 0.697]}} & \shortstack{0.573 \\ {\footnotesize [0.560, 0.586]}} & \shortstack{1.53 \\ {\footnotesize [1.49, 1.57]}} & \shortstack{$0$ \\ {\footnotesize [0, 0]}} & \shortstack{0.688 \\ {\footnotesize [0.680, 0.695]}} & \shortstack{0.575 \\ {\footnotesize [0.562, 0.587]}} & \shortstack{2.34 \\ {\footnotesize [2.10, 2.61]}} & \shortstack{$0$ \\ {\footnotesize [0, 0]}} \\
    \itc{50}  & \shortstack{0.723 \\ {\footnotesize [0.716, 0.730]}} & \shortstack{0.630 \\ {\footnotesize [0.619, 0.641]}} & \shortstack{7.50 \\ {\footnotesize [7.44, 7.55]}} & \shortstack{$0$ \\ {\footnotesize [0, 0]}} & \shortstack{0.667 \\ {\footnotesize [0.658, 0.675]}} & \shortstack{0.525 \\ {\footnotesize [0.509, 0.540]}} & \shortstack{9.42 \\ {\footnotesize [8.98, 9.89]}} & \shortstack{$0$ \\ {\footnotesize [0, 0]}} \\
    \itc{100} & \shortstack{0.726 \\ {\footnotesize [0.718, 0.733]}} & \shortstack{0.638 \\ {\footnotesize [0.627, 0.649]}} & \shortstack{15.00 \\ {\footnotesize [14.93, 15.08]}} & \shortstack{$0$ \\ {\footnotesize [0, 0]}} & \shortstack{0.678 \\ {\footnotesize [0.670, 0.686]}} & \shortstack{0.550 \\ {\footnotesize [0.535, 0.564]}} & \shortstack{18.17 \\ {\footnotesize [17.59, 18.77]}} & \shortstack{$0$ \\ {\footnotesize [0, 0]}} \\
    \itc{150} & \shortstack{0.725 \\ {\footnotesize [0.718, 0.732]}} & \shortstack{0.633 \\ {\footnotesize [0.622, 0.644]}} & \shortstack{22.56 \\ {\footnotesize [22.48, 22.65]}} & \shortstack{$0$ \\ {\footnotesize [0, 0]}} & \shortstack{0.684 \\ {\footnotesize [0.675, 0.691]}} & \shortstack{0.560 \\ {\footnotesize [0.546, 0.574]}} & \shortstack{27.31 \\ {\footnotesize [26.64, 28.02]}} & \shortstack{$0$ \\ {\footnotesize [0, 0]}} \\
    \itc{200} & \shortstack{0.725 \\ {\footnotesize [0.718, 0.732]}} & \shortstack{0.635 \\ {\footnotesize [0.623, 0.647]}} & \shortstack{30.05 \\ {\footnotesize [29.95, 30.15]}} & \shortstack{$0$ \\ {\footnotesize [0, 0]}} & \shortstack{0.685 \\ {\footnotesize [0.677, 0.693]}} & \shortstack{0.563 \\ {\footnotesize [0.549, 0.577]}} & \shortstack{36.24 \\ {\footnotesize [35.42, 37.10]}} & \shortstack{$0$ \\ {\footnotesize [0, 0]}} \\
    \itc{250} & \shortstack{0.725 \\ {\footnotesize [0.717, 0.732]}} & \shortstack{0.635 \\ {\footnotesize [0.624, 0.646]}} & \shortstack{37.63 \\ {\footnotesize [37.50, 37.76]}} & \shortstack{$0$ \\ {\footnotesize [0, 0]}} & \shortstack{0.686 \\ {\footnotesize [0.678, 0.694]}} & \shortstack{0.565 \\ {\footnotesize [0.551, 0.579]}} & \shortstack{45.26 \\ {\footnotesize [44.36, 46.19]}} & \shortstack{$0$ \\ {\footnotesize [0, 0]}} \\
    \itc{300} & \shortstack{0.726 \\ {\footnotesize [0.718, 0.733]}} & \shortstack{0.635 \\ {\footnotesize [0.624, 0.646]}} & \shortstack{45.19 \\ {\footnotesize [45.06, 45.34]}} & \shortstack{$0$ \\ {\footnotesize [0, 0]}} & \shortstack{0.689 \\ {\footnotesize [0.681, 0.696]}} & \shortstack{0.564 \\ {\footnotesize [0.549, 0.579]}} & \shortstack{54.52 \\ {\footnotesize [53.47, 55.61]}} & \shortstack{$0$ \\ {\footnotesize [0, 0]}} \\
    \itc{500} & \shortstack{0.724 \\ {\footnotesize [0.717, 0.731]}} & \shortstack{0.632 \\ {\footnotesize [0.621, 0.642]}} & \shortstack{75.50 \\ {\footnotesize [75.34, 75.66]}} & \shortstack{$0$ \\ {\footnotesize [0, 0]}} & \shortstack{0.708 \\ {\footnotesize [0.701, 0.715]}} & \shortstack{0.604 \\ {\footnotesize [0.593, 0.615]}} & \shortstack{91.03 \\ {\footnotesize [89.53, 92.57]}} & \shortstack{$0$ \\ {\footnotesize [0, 0]}} \\
    \midrule
    $\Delta$DSC$_{4,\,(50\to500)}$ & \multicolumn{4}{c}{$\textbf{+0.1\%}$} & \multicolumn{4}{c}{$+6.1\%$} \\
    $\Delta$DSC$_{35,\,(50\to500)}$ & \multicolumn{4}{c}{$\textbf{+0.3\%}$} & \multicolumn{4}{c}{$+15.0\%$} \\
    \bottomrule
    \end{tabular}}
    \end{table*}

\section{Baseline hyperparameter search}
\label{sec:supp-grid-search}

We selected the hyperparameters of the coordinate-only baseline on the validation split of the OASIS dataset (Sec.~\ref{sec:exper-setup}). The search was run with the Optuna framework \cite{akiba2019} for 15 trials over three parameters: the learning rate, drawn log-uniformly from $[1 \times 10^{-6}, 1 \times 10^{-3}]$; the number of hidden layers, drawn from 2 to 6; and the layer width, drawn from $\{64, 128, 256, 512\}$. All remaining settings, including the NCC data term and the bending-energy regularizer, were kept unchanged.

\medskip Each trial was scored on the validation pairs with the Dice score. For every pair we fit an INR from scratch to the full-resolution 3D volumes for 1,000 iterations with Adam, using coordinate batches of 10,000 points. We then applied the fitted transformation to the moving segmentation with nearest-neighbour interpolation and measured the Dice score against the fixed segmentation. The score of a trial is the mean Dice over all validation pairs, and Optuna maximizes this value. The best trial gave two hidden layers, \linebreak a width of 256, and a learning rate of $6.6 \times 10^{-5}$, which are the settings we report for the grid-searched baseline.

\section{Additional quantitative results}
\label{sec:quant}
In this section, we report the per-iteration measurements of the two coordinate-only baselines defined in Sec.~\ref{sec:exper-setup}. \linebreak Table~\ref{tab:mri-inr-meta-vs-scratch-ci} covers the INR baseline and Table~\ref{tab:mri-inr-gs-meta-vs-scratch-ci} the grid-searched variant INR$_{\mathrm{GS}}$. In both tables the left columns report the model fitted from the meta-learned initialization and the right columns the same architecture fitted from randomly initialized weights. As in Table~\ref{tab:mri-luminr-meta-vs-scratch-ci} for LUMINR, we report mean Dice score on the coarse (4-class) and the fine (35-class) labelling, cumulative fitting time, and the fraction of folding voxels over the 99 OASIS test pairs at $\alpha=0.1$, each with a 95\% confidence interval.

\section{Additional qualitative results}
\label{sec:qual}
\begin{figure*}[t]
  \centering
  \includegraphics[width=0.9\textwidth]{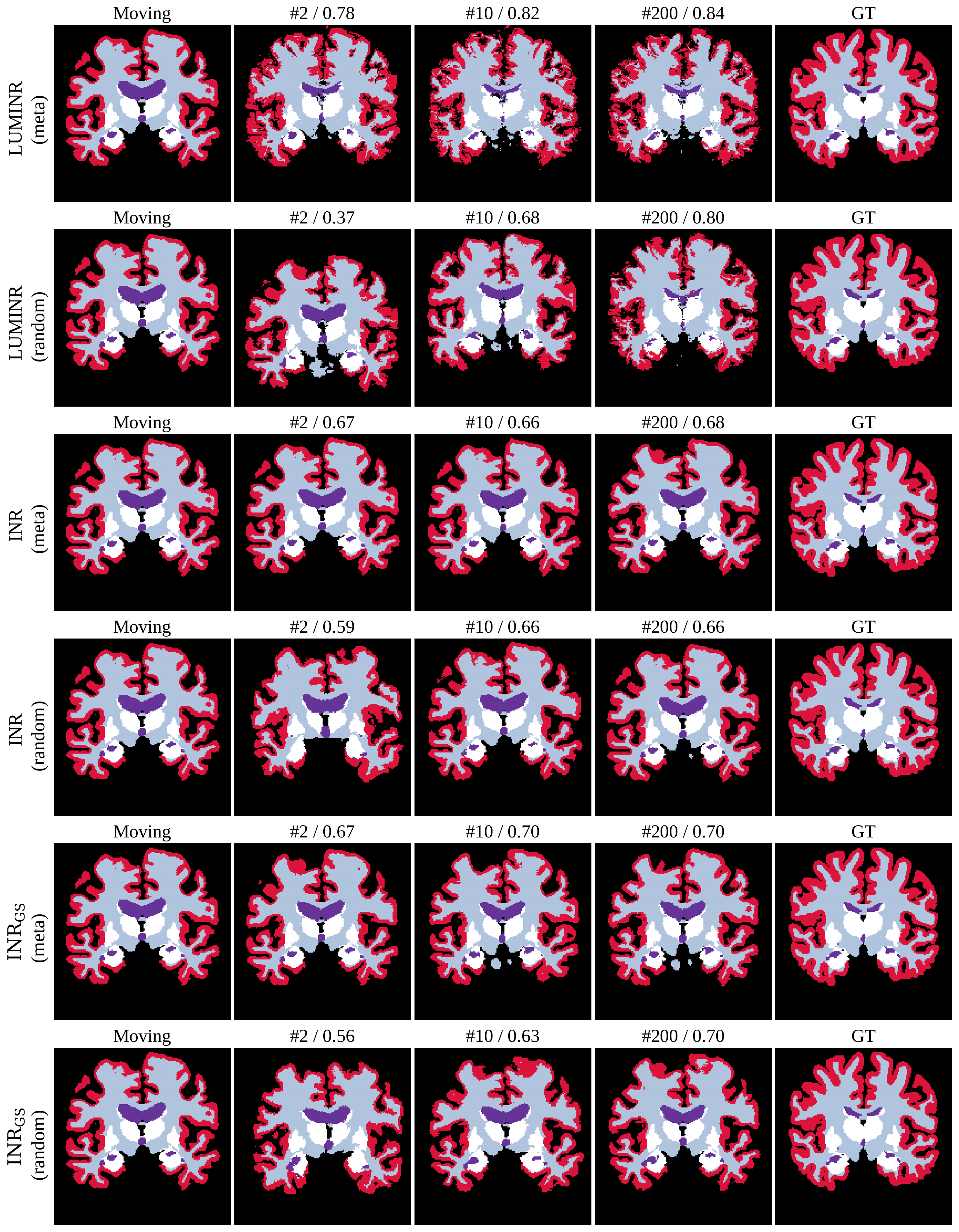}
  \caption{Comparison of segmentation performance for the LUMINR (meta), LUMINR (random), INR (meta), INR (random), INR$_{\mathrm{GS}}$ (meta), and INR$_{\mathrm{GS}}$ (random) models ($\alpha$ = 0.1) across 4 classes. We report the mean Dice score at each optimization iteration. \linebreak The meta-learned initialization leads to rapid convergence and achieves reasonable segmentation after 10 iterations.}
  \label{fig:4classes-all}
\end{figure*}

\begin{figure*}[t]
  \centering
  \includegraphics[width=0.9\textwidth]{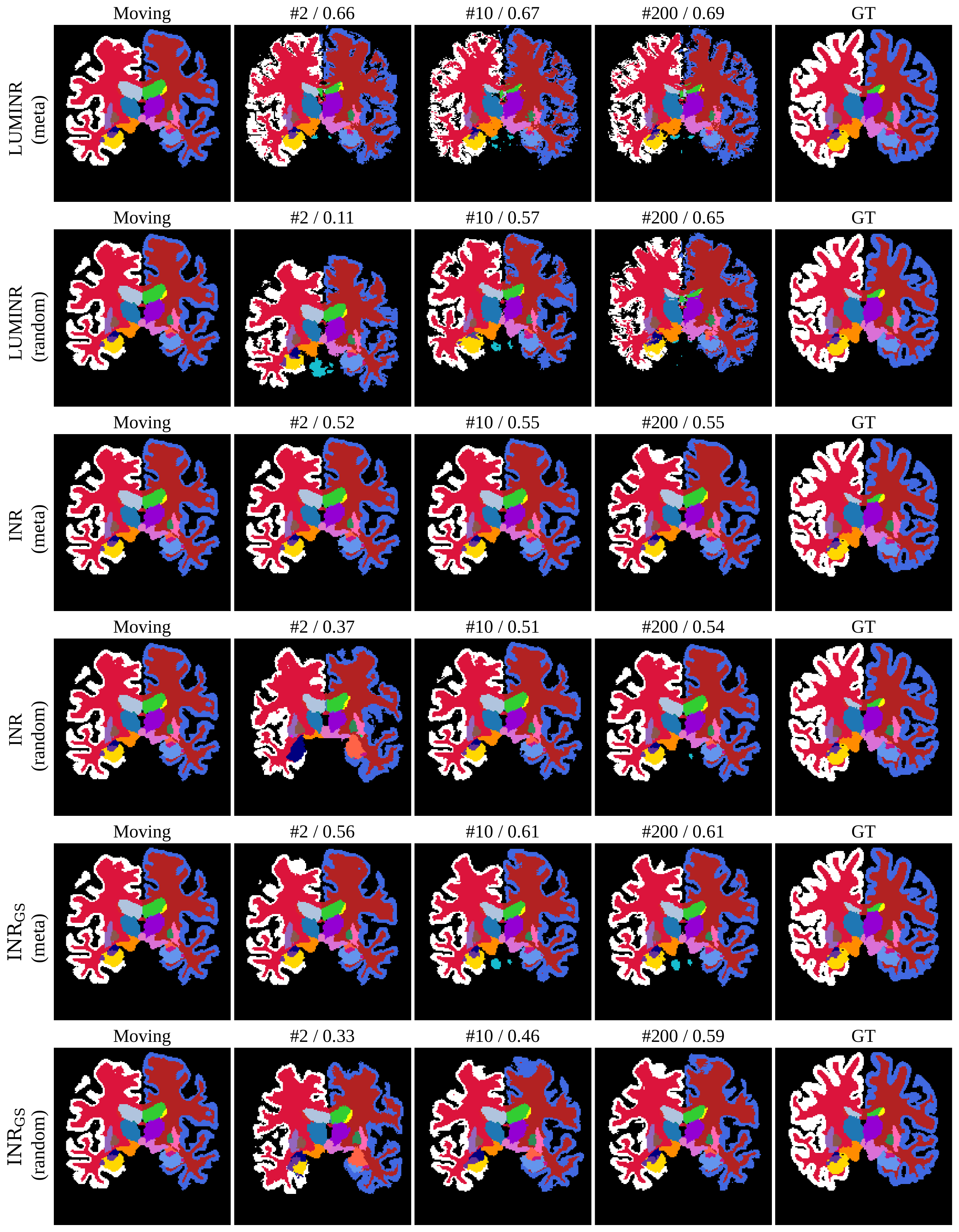}
  \caption{Comparison of segmentation performance for the LUMINR (meta), LUMINR (random), INR (meta), INR (random), INR$_{\mathrm{GS}}$ (meta), and INR$_{\mathrm{GS}}$ (random) models ($\alpha$ = 0.1) across 35 classes. We report the mean Dice score at each optimization iteration. \linebreak The meta-learned initialization leads to rapid convergence and achieves reasonable segmentation after 10 iterations.}
  \label{fig:35classes-all}
\end{figure*}

We also show the warped segmentation masks of all six configurations defined in Sec.~\ref{sec:exper-setup} on one test pair. Fig.~\ref{fig:4classes-all} uses the coarse 4-class labelling and Fig.~\ref{fig:35classes-all} the fine 35-class one, and both share a layout of six rows and five columns. A row is one architecture, LUMINR, INR, or INR$_{\mathrm{GS}}$, fitted from the meta-learned initialization, and the row below it is the same architecture fitted from randomly initialized weights. The first column is the moving segmentation and the last one the ground truth of the fixed volume; the three columns between them are fitting iterations 2, 10, and 200 at $\alpha=0.1$, each headed by its iteration number and by the Dice score of the warped volume. Every panel shows the same central axial slice, so the masks are a 2D section through the volumes the Dice scores are measured on.

\begin{table*}[t]
    \centering
    \caption{Mean Dice score, cumulative fitting time, and fraction of folding voxels across fitting iterations for the baseline INR$_{\mathrm{GS}}$ (meta) and INR$_{\mathrm{GS}}$ (random) models. DSC$_4$ and DSC$_{35}$ are Dice on the coarse (4-class) and fine (35-class) labellings respectively; fitting time and folding are independent of the labelling and are therefore reported once. Each entry is the mean over all test pairs with its 95\% confidence interval in brackets. Both models are the same architecture: the left one is fitted from the meta-learned initialization, the right one from \linebreak a random initialization.}
    \label{tab:mri-inr-gs-meta-vs-scratch-ci}
    \footnotesize
    \newcommand{\itc}[1]{\raisebox{0.5\ht\strutbox}[0pt][0pt]{#1}}
    \setlength{\tabcolsep}{4pt}
    \resizebox{\textwidth}{!}{%
    \begin{tabular}{@{}ccccccccc@{}}
    \toprule
     & \multicolumn{4}{c}{INR$_{\mathrm{GS}}$ (meta)} & \multicolumn{4}{c}{INR$_{\mathrm{GS}}$ (random)} \\
    \cmidrule(lr){2-5} \cmidrule(lr){6-9}
    Iter. & DSC$_4$ & DSC$_{35}$ & $t$ & det($J$)$<$0\,$\times10^{-5}$ & DSC$_4$ & DSC$_{35}$ & $t$ & det($J$)$<$0\,$\times10^{-5}$ \\
          & [-] & [-] & [s] & [\%] & [-] & [-] & [s] & [\%] \\
    \midrule
    \itc{2}   & \shortstack{0.701 \\ {\footnotesize [0.694, 0.709]}} & \shortstack{0.598 \\ {\footnotesize [0.586, 0.610]}} & \shortstack{0.17 \\ {\footnotesize [0.14, 0.22]}} & \shortstack{$0$ \\ {\footnotesize [0, 0]}} & \shortstack{0.603 \\ {\footnotesize [0.594, 0.612]}} & \shortstack{0.399 \\ {\footnotesize [0.383, 0.415]}} & \shortstack{0.15 \\ {\footnotesize [0.12, 0.18]}} & \shortstack{$0$ \\ {\footnotesize [0, 0]}} \\
    \itc{10}  & \shortstack{0.730 \\ {\footnotesize [0.723, 0.737]}} & \shortstack{0.642 \\ {\footnotesize [0.631, 0.653]}} & \shortstack{0.74 \\ {\footnotesize [0.68, 0.83]}} & \shortstack{$0$ \\ {\footnotesize [0, 0]}} & \shortstack{0.671 \\ {\footnotesize [0.663, 0.679]}} & \shortstack{0.530 \\ {\footnotesize [0.516, 0.544]}} & \shortstack{0.66 \\ {\footnotesize [0.62, 0.72]}} & \shortstack{$0$ \\ {\footnotesize [0, 0]}} \\
    \itc{50}  & \shortstack{0.736 \\ {\footnotesize [0.729, 0.743]}} & \shortstack{0.655 \\ {\footnotesize [0.644, 0.666]}} & \shortstack{3.61 \\ {\footnotesize [3.49, 3.75]}} & \shortstack{$0$ \\ {\footnotesize [0, 0]}} & \shortstack{0.709 \\ {\footnotesize [0.701, 0.716]}} & \shortstack{0.611 \\ {\footnotesize [0.599, 0.623]}} & \shortstack{3.22 \\ {\footnotesize [3.14, 3.32]}} & \shortstack{$0$ \\ {\footnotesize [0, 0]}} \\
    \itc{100} & \shortstack{0.737 \\ {\footnotesize [0.730, 0.744]}} & \shortstack{0.655 \\ {\footnotesize [0.644, 0.665]}} & \shortstack{7.35 \\ {\footnotesize [7.17, 7.54]}} & \shortstack{$0$ \\ {\footnotesize [0, 0]}} & \shortstack{0.717 \\ {\footnotesize [0.709, 0.724]}} & \shortstack{0.624 \\ {\footnotesize [0.612, 0.635]}} & \shortstack{6.36 \\ {\footnotesize [6.26, 6.48]}} & \shortstack{$0$ \\ {\footnotesize [0, 0]}} \\
    \itc{150} & \shortstack{0.738 \\ {\footnotesize [0.730, 0.744]}} & \shortstack{0.654 \\ {\footnotesize [0.642, 0.664]}} & \shortstack{11.05 \\ {\footnotesize [10.83, 11.29]}} & \shortstack{$0$ \\ {\footnotesize [0, 0]}} & \shortstack{0.721 \\ {\footnotesize [0.714, 0.728]}} & \shortstack{0.632 \\ {\footnotesize [0.621, 0.642]}} & \shortstack{9.55 \\ {\footnotesize [9.41, 9.70]}} & \shortstack{$0$ \\ {\footnotesize [0, 0]}} \\
    \itc{200} & \shortstack{0.738 \\ {\footnotesize [0.731, 0.745]}} & \shortstack{0.655 \\ {\footnotesize [0.644, 0.666]}} & \shortstack{14.71 \\ {\footnotesize [14.46, 14.98]}} & \shortstack{$0$ \\ {\footnotesize [0, 0]}} & \shortstack{0.726 \\ {\footnotesize [0.719, 0.733]}} & \shortstack{0.639 \\ {\footnotesize [0.628, 0.648]}} & \shortstack{12.74 \\ {\footnotesize [12.58, 12.91]}} & \shortstack{$0$ \\ {\footnotesize [0, 0]}} \\
    \itc{250} & \shortstack{0.738 \\ {\footnotesize [0.731, 0.745]}} & \shortstack{0.656 \\ {\footnotesize [0.645, 0.666]}} & \shortstack{18.40 \\ {\footnotesize [18.12, 18.69]}} & \shortstack{$0$ \\ {\footnotesize [0, 0]}} & \shortstack{0.728 \\ {\footnotesize [0.721, 0.735]}} & \shortstack{0.641 \\ {\footnotesize [0.631, 0.652]}} & \shortstack{15.91 \\ {\footnotesize [15.72, 16.10]}} & \shortstack{$0$ \\ {\footnotesize [0, 0]}} \\
    \itc{300} & \shortstack{0.738 \\ {\footnotesize [0.731, 0.745]}} & \shortstack{0.654 \\ {\footnotesize [0.643, 0.665]}} & \shortstack{22.05 \\ {\footnotesize [21.76, 22.36]}} & \shortstack{$0$ \\ {\footnotesize [0, 0]}} & \shortstack{0.731 \\ {\footnotesize [0.724, 0.738]}} & \shortstack{0.647 \\ {\footnotesize [0.636, 0.658]}} & \shortstack{19.07 \\ {\footnotesize [18.86, 19.28]}} & \shortstack{$0$ \\ {\footnotesize [0, 0]}} \\
    \itc{500} & \shortstack{0.739 \\ {\footnotesize [0.731, 0.745]}} & \shortstack{0.655 \\ {\footnotesize [0.644, 0.666]}} & \shortstack{36.69 \\ {\footnotesize [36.32, 37.08]}} & \shortstack{$0$ \\ {\footnotesize [0, 0]}} & \shortstack{0.734 \\ {\footnotesize [0.727, 0.741]}} & \shortstack{0.651 \\ {\footnotesize [0.640, 0.661]}} & \shortstack{31.76 \\ {\footnotesize [31.47, 32.06]}} & \shortstack{$0$ \\ {\footnotesize [0, 0]}} \\
    \midrule
    $\Delta$DSC$_{4,\,(50\to500)}$ & \multicolumn{4}{c}{$\textbf{+0.4\%}$} & \multicolumn{4}{c}{$+3.5\%$} \\
    $\Delta$DSC$_{35,\,(50\to500)}$ & \multicolumn{4}{c}{$\textbf{+0.0\%}$} & \multicolumn{4}{c}{$+6.5\%$} \\
    \bottomrule
    \end{tabular}}
    \end{table*}

%% file: main.bib
@String(CVPR  = {CVPR})

@String(ICCV  = {ICCV})

@String(NIPS  = {NeurIPS})

@String(ICLR  = {ICLR})

@String(MICCAI = {International Conference on Medical Image Computing and Computer-Assisted Intervention (MICCAI)})

@String(MIDL   = {Medical Imaging with Deep Learning (MIDL)})

@String(ICML   = {International Conference on Machine Learning (ICML)})

@String(TMI    = {IEEE Transactions on Medical Imaging})

@inproceedings{akiba2019,
  title     = {{O}ptuna: A Next-Generation Hyperparameter Optimization Framework},
  author    = {Akiba, Takuya and Sano, Shotaro and Yanase, Toshihiko and Ohta, Takeru and Koyama, Masanori},
  booktitle = {Proceedings of the 25th {ACM} {SIGKDD} International Conference on Knowledge Discovery \& Data Mining},
  pages     = {2623--2631},
  year      = {2019},
  doi       = {10.1145/3292500.3330701}
}

@article{baum2023,
  title   = {Meta-Learning Initializations for Interactive Medical Image Registration},
  author  = {Baum, Zachary M. C. and Hu, Yipeng and Barratt, Dean C.},
  journal = {IEEE Transactions on Medical Imaging},
  volume  = {42},
  number  = {3},
  pages   = {823--833},
  year    = {2023},
  doi     = {10.1109/TMI.2022.3218147}
}

@article{balakrishnan2019,
  title   = {VoxelMorph: A Learning Framework for Deformable Medical Image Registration},
  author  = {Balakrishnan, Guha and Zhao, Amy and Sabuncu, Mert R. and Guttag, John and Dalca, Adrian V.},
  journal = {IEEE Transactions on Medical Imaging},
  volume  = {38},
  number  = {8},
  pages   = {1788--1800},
  year    = {2019},
  doi     = {10.1109/TMI.2019.2897538}
}

@inproceedings{dupont2022,
  author    = {Emilien Dupont and Hyunjik Kim and S. M. Ali Eslami and Danilo J. Rezende and Dan Rosenbaum},
  title     = {From data to functa: Your data point is a function and you can treat it like one},
  booktitle = ICML,
  year      = {2022},
}

@inproceedings{finn2017,
  author    = {Chelsea Finn and Pieter Abbeel and Sergey Levine},
  title     = {Model-agnostic meta-learning for fast adaptation of deep networks},
  booktitle = ICML,
  pages     = {1126--1135},
  year      = {2017},
}

@inproceedings{friedrich2026,
  author    = {Paul Friedrich and Florentin Bieder and Julian McGinnis and Julia Wolleb and Daniel Rueckert and Philippe C. Cattin},
  title     = {{MedFuncta}: A unified framework for learning efficient medical neural fields},
  booktitle = MIDL,
  year      = {2026},
}

@inproceedings{harten2024,
  author    = {van Harten, L. and van Herten, R. L. M. and Isgum, I.},
  title     = {{REINDIR}: Repeated embedding infusion for neural deformable image registration},
  booktitle = MIDL,
  pages     = {577--595},
  year      = 2024
}

@article{hering2022,
  author  = {Alessa Hering and Lasse Hansen and Tony C. W. Mok and others},
  title   = {{Learn2Reg}: Comprehensive multi-task medical image registration challenge, dataset and evaluation in the era of deep learning},
  journal = TMI,
  volume  = {42},
  number  = {3},
  pages   = {697--712},
  year    = {2023},
}

@article{hoopes2022,
  author  = {Andrew Hoopes and Malte Hoffmann and Douglas N. Greve and Bruce Fischl and John Guttag and Adrian V. Dalca},
  title   = {Learning the effect of registration hyperparameters with {HyperMorph}},
  journal = {Machine Learning for Biomedical Imaging},
  volume  = {1},
  pages   = {1--30},
  year    = {2022},
}

@article{khan2023,
  author  = {A. Khan and Z. Rauf and A. R. Khan and S. Rathore and S. H. Khan and N. S. Shah and U. Farooq and H. Asif and A. Asif and U. Zahoora and R. U. Khalil and S. Qamar and U. H. Asif and F. B. Khan and A. Majid and J. Gwak},
  title   = {A recent survey of vision transformers for medical image segmentation},
  journal = {IEEE Access},
  volume  = {13},
  pages   = {191824--191849},
  year    = {2025},
}

@inproceedings{kingma2014,
  author    = {Diederik P. Kingma and Jimmy Ba},
  title     = {{Adam}: A method for stochastic optimization},
  booktitle = ICLR,
  year      = {2015},
}

@article{marcus2007,
  author  = {Daniel S. Marcus and Tracy H. Wang and Jamie Parker and John G. Csernansky and John C. Morris and Randy L. Buckner},
  title   = {Open {Access} {Series} of {Imaging} {Studies} ({OASIS}): Cross-sectional {MRI} data in young, middle aged, nondemented, and demented older adults},
  journal = {Journal of Cognitive Neuroscience},
  volume  = {19},
  number  = {9},
  pages   = {1498--1507},
  year    = {2007},
}

@article{nichol2018,
  author  = {Alex Nichol and Joshua Achiam and John Schulman},
  title   = {On first-order meta-learning algorithms},
  journal = {arXiv preprint arXiv:1803.02999},
  year    = {2018},
}

@inproceedings{reed2021,
  author    = {Albert W. Reed and Hyojin Kim and Rushil Anirudh and K. Aditya Mohan and Kyle Champley and Jingu Kang and Suren Jayasuriya},
  title     = {Dynamic {CT} reconstruction from limited views with implicit neural representations and parametric motion fields},
  booktitle = ICCV,
  pages     = {2258--2268},
  year      = {2021},
}

@inproceedings{ronneberger2015,
  author    = {Olaf Ronneberger and Philipp Fischer and Thomas Brox},
  title     = {{U-Net}: Convolutional networks for biomedical image segmentation},
  booktitle = MICCAI,
  pages     = {234--241},
  year      = {2015},
}

@article{rueckert1999,
  author  = {Daniel Rueckert and Luke I. Sonoda and Carmel Hayes and Derek L. G. Hill and Martin O. Leach and David J. Hawkes},
  title   = {Nonrigid registration using free-form deformations: Application to breast {MR} images},
  journal = TMI,
  volume  = {18},
  number  = {8},
  pages   = {712--721},
  year    = {1999},
  doi     = {10.1109/42.796284},
}

@article{shen2022,
  author  = {Liyue Shen and John Pauly and Lei Xing},
  title   = {{NeRP}: Implicit neural representation learning with prior embedding for sparsely sampled image reconstruction},
  journal = {IEEE Transactions on Neural Networks and Learning Systems},
  year    = {2022},
}

@inproceedings{sitzmann2020,
  author    = {Vincent Sitzmann and Julien N. P. Martel and Alexander W. Bergman and David B. Lindell and Gordon Wetzstein},
  title     = {Implicit neural representations with periodic activation functions},
  booktitle = NIPS,
  pages     = {7462--7473},
  year      = {2020},
}

@inproceedings{tancik2021,
  author    = {Matthew Tancik and Ben Mildenhall and Terrance Wang and Divi Schmidt and Pratul P. Srinivasan and Jonathan T. Barron and Ren Ng},
  title     = {Learned initializations for optimizing coordinate-based neural representations},
  booktitle = CVPR,
  year      = {2021},
}

@inproceedings{vyas2025,
  author    = {Kushal Vyas and Ashok Veeraraghavan and Guha Balakrishnan},
  title     = {Fit pixels, get labels: Meta-learned implicit networks for image segmentation},
  booktitle = MICCAI,
  pages     = {194--203},
  year      = {2025},
}

@inproceedings{vyas2024,
  author    = {Kushal Kumar Vyas and Ahmed Imtiaz Humayun and Aniket Dashpute and Richard Baraniuk and Ashok Veeraraghavan and Guha Balakrishnan},
  title     = {Learning transferable features for implicit neural representations},
  booktitle = NIPS,
  year      = {2024},
}

@inproceedings{wiesner2022,
  author    = {David Wiesner and Julian Suk and Sven Dummer and David Svoboda and Jelmer M. Wolterink},
  title     = {Implicit neural representations for generative modeling of living cell shapes},
  booktitle = MICCAI,
  pages     = {58--67},
  year      = {2022},
}

@inproceedings{wolterink2022,
  author    = {Jelmer M. Wolterink and Jesse C. Zwienenberg and Christoph Brune},
  title     = {Implicit neural representations for deformable image registration},
  booktitle = MIDL,
  year      = {2022},
}
